\documentclass[12pt,a4paper]{article}
\usepackage{amsmath,amssymb}
\usepackage[sort]{cite}
\usepackage{graphicx}
\newcommand{\ind}[2]{^{#1}_{\text{#2}}}

\def\mmu{m_{\mu}}
\def\DP{\Delta\Pi}
\def\iep{i\varepsilon}
\def\arctanh{{\rm arctanh}}
\newcommand{\Li}[1]{{\rm Li}_{#1}}
\newcommand{\amu}[1]{a^{\text{HVP}(#1)}_{\mu}}
\newcommand{\APF}[1]{A_{0}^{(#1)}}
\newcommand{\KG}[2]{K_{#1}^{(#2)}}
\newcommand{\KGt}[2]{\tilde{K}_{#1}^{(#2)}}

\begin{document}

\begin{center}

{\Large\bf Timelike and spacelike kernel functions for \\
the hadronic vacuum polarization contribution \\ 
to the muon anomalous magnetic moment

}

\vskip10mm

{\large A.V.~Nesterenko}

\vskip7.5mm

{\small\it Bogoliubov Laboratory of Theoretical Physics,
Joint Institute for Nuclear Research,\\
Dubna, 141980, Russian Federation}

\end{center}

\vskip5mm

\noindent
\centerline{\bf Abstract}

\vskip2.5mm

\centerline{\parbox[t]{150mm}{%
The complete set of relations, which mutually express the spacelike and 
timelike kernel functions for the hadronic vacuum polarization contribution 
to the muon anomalous magnetic moment~$a^{\text{HVP}}_{\mu}$ in terms of 
each other, is obtained. By~making use of the derived relations the 
explicit expression for the next--to--leading order spacelike kernel 
function, which enters the representation for~$a^{\text{HVP}}_{\mu}$ 
involving the hadronic vacuum polarization function, is obtained. 
The corresponding next--to--leading order spacelike kernel function, 
which appears in the representation for~$a^{\text{HVP}}_{\mu}$ involving 
the Adler function, is calculated numerically. The~obtained results 
can be employed in the assessments of the hadronic vacuum polarization 
contribution to the muon anomalous magnetic moment in the framework 
of the spacelike methods, such as lattice studies, MUonE project, 
and others.
\\[2.5mm]
\textbf{Keywords:}~\parbox[t]{127mm}{%
muon anomalous magnetic moment, hadronic vacuum polarization
contributions, kernel functions, lattice~QCD}%
}}

\vskip12mm

\section{Introduction}
\label{Sect:Intro}

The hadronic contribution to the muon anomalous magnetic moment 
$a_{\mu} = (g_{\mu}-2)/2$ represents one of the long--standing 
challenging issues of elementary particle physics, which engages 
the entire pattern of interactions within the Standard Model.
The~experimental measurements~\cite{BNL06, FNAL21} and theoretical 
evaluations (see a recent comprehensive review~\cite{WP20}, which 
is mainly based 
on Refs.~\cite{Davier:2010nc, Davier:2017zfy, Keshavarzi:2018mgv,
Colangelo:2018mtw, Hoferichter:2019mqg, Davier:2019can, Keshavarzi:2019abf,
Kurz:2014wya, FermilabLattice:2017wgj, Budapest-Marseille-Wuppertal:2017okr,
RBC:2018dos, Giusti:2019xct, Shintani:2019wai, FermilabLattice:2019ugu,
Gerardin:2019rua, Aubin:2019usy, Giusti:2019hkz, Melnikov:2003xd,
Masjuan:2017tvw, Colangelo:2017fiz, Hoferichter:2018kwz, Gerardin:2019vio,
Bijnens:2019ghy, Colangelo:2019uex, Pauk:2014rta, Danilkin:2016hnh,
Jegerlehner:2017gek, Knecht:2018sci, Eichmann:2019bqf, Roig:2019reh,
Colangelo:2014qya, Blum:2019ugy, Aoyama:2012wk, Aoyama:2019ryr,
Czarnecki:2002nt, Gnendiger:2013pva}) of this quantity have achieved 
an impressive 
accuracy, and the remaining discrepancy of the order of a few standard 
deviations between them may be an evidence for the existence of a new 
fundamental physics beyond the Standard Model. The~uncertainty of 
theoretical estimation of~$a_{\mu}$ is largely dominated by the 
hadronic contribution, which involves the tangled dynamics of colored
fields in the infrared domain inaccessible within perturbation theory.

There are basically two approaches to the theoretical assessment
of the hadronic vacuum polarization contributions to the muon
anomalous magnetic moment~$a^{\text{HVP}}_{\mu}$. 
Specifically, in the framework
of the first (``spacelike'') approach~$a^{\text{HVP}}_{\mu}$ is commonly
represented as the integral of the hadronic vacuum polarization
function~$\bar\Pi(Q^2)$ [or~the related Adler function~$D(Q^2)$]
with corresponding kernel functions~$K_{\Pi}(Q^2)$ [or~$K_{D}(Q^2)$]
over the entire kinematic interval. Here the perturbative results
for the involved functions~$\bar\Pi(Q^2)$ and~$D(Q^2)$ have to be
supplemented with the relevant nonperturbative inputs. The latter can
be provided~by,~e.g., lattice simulations~\cite{Lattice1, Lattice2, BMW21}
(which, being capable of delivering valuable insights into the
underlying mechanisms, have a large scientific potential), highly
anticipated MUonE measurements~\cite{MUonE1, MUonE2, MUonE3}, 
and other methods.
Alternatively, in the framework of the second (``timelike'')
approach~$a^{\text{HVP}}_{\mu}$ can also be represented as 
the integral of the
function~$R(s)$ with respective kernel functions~$K_{R}(s)$.
Here the perturbative results for the function~$R(s)$ are usually
complemented~by the low--energy experimental data on 
the~\mbox{$R$--ratio} of electron--positron annihilation into
hadrons, that constitutes the data--driven method of evaluation
of~$a^{\text{HVP}}_{\mu}$. In~turn, the ``spacelike'' and 
``timelike'' kernel
functions can be calculated within various techniques, such~as
the mass operator approach~\cite{Schwinger, Milton74a, Milton74b},
the hyperspherical approach~\cite{HS1, HS2, HS3, HS4}, 
the dispersive method~\cite{DispMeth1, DispMeth2, DispMeth3, DispMeth4,
DispMeth5}, and the asymptotic expansion method~\cite{Smirnov12,
Krause96, Kurz:2014wya}. The~``timelike'' kernel functions have been
extensively studied over the past decades, whereas 
the corresponding ``spacelike'' kernel functions remain largely
unavailable.

The primary objective of this paper is to derive the complete set of 
relations, which mutually express the~``spacelike'' and~``timelike'' 
kernel functions~$K_{\Pi}(Q^2)$, $K_{D}(Q^2)$, and~$K_{R}(s)$ in terms 
of each other, and to calculate the explicit expression for the
next--to--leading order ``spacelike'' kernel function~$\KG{\Pi}{3a}(Q^2)$ 
by making use of the obtained relations.

The layout of the paper is as follows. Section~\ref{Sect:Methods} 
recaps the essentials of the dispersion relations for the hadronic 
vacuum polarization function~$\bar\Pi(Q^2)$, the Adler function~$D(Q^2)$, 
and~the function~$R(s)$, and expounds the basics of the hadronic 
vacuum polarization contributions to the muon anomalous magnetic
moment. In~Sect.~\ref{Sect:Results} the complete set of relations, 
which mutually express the kernel functions~$K_{\Pi}(Q^2)$, $K_{D}(Q^2)$, 
and~$K_{R}(s)$ in terms of each other, is~obtained, and the explicit 
expression for the ``spacelike'' kernel function~$\KG{\Pi}{3a}(Q^2)$ 
is~calculated. Section~\ref{Sect:Concl} summarizes the basic results.
The ``timelike'' kernel function~$\KG{R}{3a}(s)$ is given in 
the App.~\ref{Sect:KR3aExpl}.

\section{Methods}
\label{Sect:Methods}

\subsection{General dispersion relations}
\label{Sect:GDR}

Let us begin by briefly elucidating the essentials of dispersion 
relations for the hadronic vacuum polarization function~$\Pi(q^2)$, 
the Adler function~$D(Q^2)$, and the function~$R(s)$ (the~detailed 
description of this issue can be found in,~e.g., Chap.~1 of 
Ref.~\cite{Book} and references therein). The theoretical 
exploration of a certain class of the strong interaction 
processes is primarily based on the hadronic vacuum polarization 
function~$\Pi(q^2)$, which is defined as the scalar part of the 
hadronic vacuum polarization tensor
\begin{equation}
\label{P_Def}
\Pi_{\mu\nu}(q^2) = i\!\int\!d^4x\,e^{i q x} \bigl\langle 0 \bigl|\,
T\!\left\{J_{\mu}(x)\, J_{\nu}(0)\right\} \bigr| 0 \bigr\rangle =
\frac{i}{12\pi^2} (q_{\mu}q_{\nu} - g_{\mu\nu}q^2) \Pi(q^2).
\end{equation}
As discussed in, e.g., Ref.~\cite{Feynman}, the 
function~$\Pi(q^2)$~(\ref{P_Def}) has the
only cut along the positive semiaxis of real~$q^2$ starting at the
hadronic production threshold~$q^2 \ge s_{0}$, that leads~to
\begin{equation}
\label{PDisp}
\DP(q^{2},q_{0}^{2}) = 
(q^2-q_{0}^{2})\int\limits_{s_{0}}^{\infty}
\frac{R(\sigma)}{(\sigma-q^2)(\sigma-q_0^2)}\, d\sigma,
\end{equation}
where
\begin{equation}
\label{PSubDef}
\DP(q^{2},q_{0}^{2}) = \Pi(q^2) - \Pi(q_0^2),
\qquad
\Pi(0)=0,
\qquad
\DP(0,-p^2)=-\Pi(-p^2)=\bar\Pi(p^2),
\end{equation}
and
\begin{equation}
\label{RDefP}
R(s) = \frac{1}{2 \pi i} \lim_{\varepsilon \to 0_{+}}
\Bigl[\Pi(s + i \varepsilon) - \Pi(s - i \varepsilon)\Bigr].
\end{equation}
The function~$R(s)$~(\ref{RDefP}) is commonly identified with the 
so--called $R$--ratio of electron--positron annihilation into hadrons 
\begin{equation}
\label{RDefExp}
R(s) =
\frac{\sigma(e^{+}e^{-} \to \text{hadrons}; s)}{\sigma(e^{+}e^{-} \to
\mu^{+}\mu^{-}; s)}\,,
\end{equation}
where \mbox{$s=q^2>0$} is the timelike
kinematic variable, namely, the center--of--mass energy squared.
At~the same time, in~practical applications it proves to be 
convenient to deal with the Adler function~\cite{Adler}
\begin{equation}
\label{GDR_DP}
D(Q^2) = -\,\frac{d\, \Pi(-Q^2)}{d \ln Q^2},
\end{equation}
where~$Q^2=-q^2>0$ stands for the spacelike kinematic variable. The
corresponding dispersion relation~\cite{Adler}
\begin{equation}
\label{GDR_DR}
D(Q^2) = 
Q^2 \int\limits_{s_{0}}^{\infty}
\frac{R(\sigma)}{(\sigma+Q^2)^2}\, d\sigma
\end{equation}
immediately follows from Eqs.~(\ref{PDisp}) and~(\ref{GDR_DP}).

\subsection{Hadronic vacuum polarization contributions to~$a_{\mu}$}
\label{AmuHVP}

\begin{figure}[t]
\centerline{\includegraphics[height=50mm,clip]{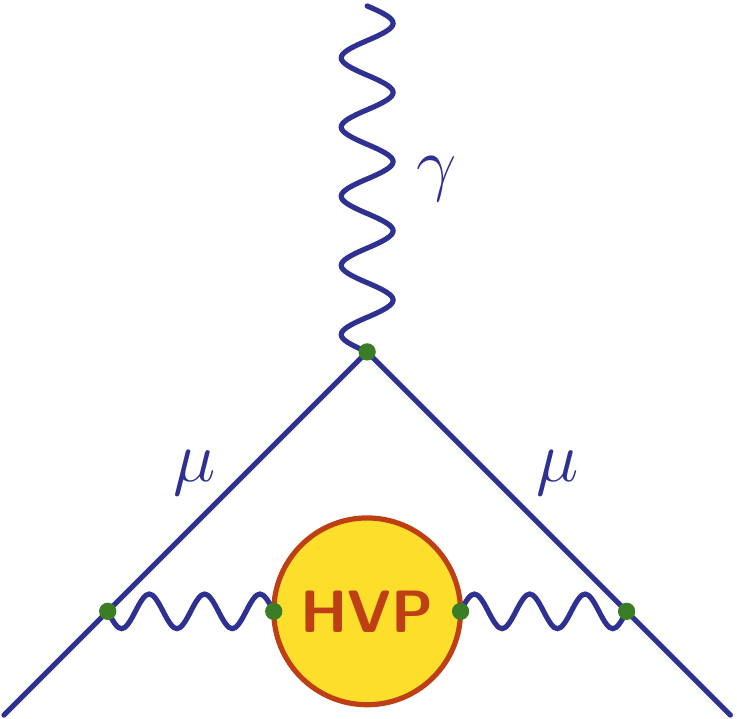}}
\caption{The leading--order hadronic vacuum polarization 
contribution to the muon anomalous magnetic moment~(\ref{Amu2Def}).}
\label{Plot:Amu2}
\end{figure}

In the framework of the ``timelike'' (or~data--driven) method of
assessment of the hadronic vacuum polarization contributions to
the muon anomalous magnetic moment the latter is commonly represented in
terms of the $R$--ratio of electron--positron annihilation into
hadrons~(\ref{RDefP}). In~the leading order of perturbation theory
(namely, in the second order in the electromagnetic coupling)
the corresponding contribution is given by the diagram displayed
in Fig.~\ref{Plot:Amu2}, that yields~\cite{BKK56, BM61, KO67}
\begin{equation}
\label{Amu2Def}
\amu{2} = \frac{1}{3} \Bigl(\frac{\alpha}{\pi}\Bigr)^{\!2}
\!\int\limits_{s_{0}}^{\infty}
\frac{G_{2}(s)}{s} R(s) ds,
\end{equation}
where
\begin{equation}
\label{K2RInt}
G_{2}(s) = \int\limits_{0}^{1}\!
\frac{x^2 (1-x)}{x^2 + (1-x) s/\mmu^2}\,dx
\end{equation}
and~$s=q^2 \ge 0$ denotes the timelike kinematic variable.
The kernel function~$G_{2}(s)$~(\ref{K2RInt}) can also be
represented in explicit form~\cite{BKK56, D6263, BdR67, LdR68} and
the expression appropriate for the practical 
applications reads
\begin{equation}
\label{K2RExpl}
G_{2}(s)  = \frac{1}{2} + 4\eta\Bigl[(2\eta-1)\ln(4\eta)-1\Bigr]
-2\Bigl[2(2\eta-1)^2-1\Bigr]
\frac{A(\eta)}{\psi(\eta)},
\end{equation}
where
\begin{equation}
\label{DefAux1}
\psi(\eta) = \frac{\sqrt{\eta-1}}{\sqrt{\eta}},
\qquad
A(\eta) = \arctanh\Bigl[\psi(\eta)\Bigr],
\qquad
\eta=\frac{s}{4\mmu^2}.
\end{equation}

Factually, the specific form of the kernel 
function~$G_{2}(s)$~(\ref{K2RInt})
makes it possible to express the leading--order
contribution~(\ref{Amu2Def})
in terms of the ``spacelike'' hadronic vacuum polarization 
function~$\bar\Pi(Q^2)$ [Eqs.~(\ref{PDisp}),~(\ref{PSubDef})] 
and the Adler function~$D(Q^2)$~(\ref{GDR_DP}),
but only in this particular case
(a~discussion of this issue can 
be found in, e.g.,~Ref.~\cite{EdR17}). Namely, Eqs.~(\ref{Amu2Def}), 
(\ref{K2RInt}),~(\ref{PDisp}), and~(\ref{PSubDef}) can be 
reduced~to~\cite{LPdR71}
\begin{align}
\label{Amu2Px}
\amu{2} & = \frac{1}{3} \Bigl(\frac{\alpha}{\pi}\Bigr)^{\!2}
\!\int\limits_{0}^{1}\! d x (1-x)
\!\!\int\limits_{s_{0}}^{\infty}\!
\frac{d s}{s} 
\frac{\mmu^2 x^2 (1-x)^{-1}}{s + \mmu^2 x^2 (1-x)^{-1}} R(s)
= \nonumber \\ &
= \frac{1}{3} \Bigl(\frac{\alpha}{\pi}\Bigr)^{\!2}
\!\int\limits_{0}^{1}\! (1-x) 
\bar\Pi\biggl(\!\mmu^2\frac{x^2}{1-x}\biggr) d x,
\end{align}
where~$\bar\Pi(Q^2)=\DP(0,-Q^2)$ and~$Q^2 = -q^2 \ge 0$
stands for the spacelike kinematic variable.
In~turn, Eq.~(\ref{Amu2Px}) can also be represented in terms 
of the Adler function~(\ref{GDR_DP}) by making use of the
integration by parts, that eventually yields~\cite{Knecht2004, EdR17}
\begin{equation}
\label{Amu2Dx}
\amu{2} = \frac{1}{3} \Bigl(\frac{\alpha}{\pi}\Bigr)^{\!2}
\!\int\limits_{0}^{1}\! (1-x) \biggl(\!1-\frac{x}{2}\biggr)
D\biggl(\!\mmu^2\frac{x^2}{1-x}\biggr) \frac{d x}{x}.
\end{equation}
It~is necessary to emphasize here that this way of the
derivation of the ``spacelike'' expressions~(\ref{Amu2Px}) 
and~(\ref{Amu2Dx}) 
from the ``timelike'' one~(\ref{Amu2Def}) entirely relies on the
particular form of the leading--order kernel 
function~$G_{2}(s)$~(\ref{K2RInt})
and cannot be performed in any other case.

\bigskip

\begin{figure}[t]
\centerline{%
\begin{tabular}{lcr}
\includegraphics[height=50mm,clip]{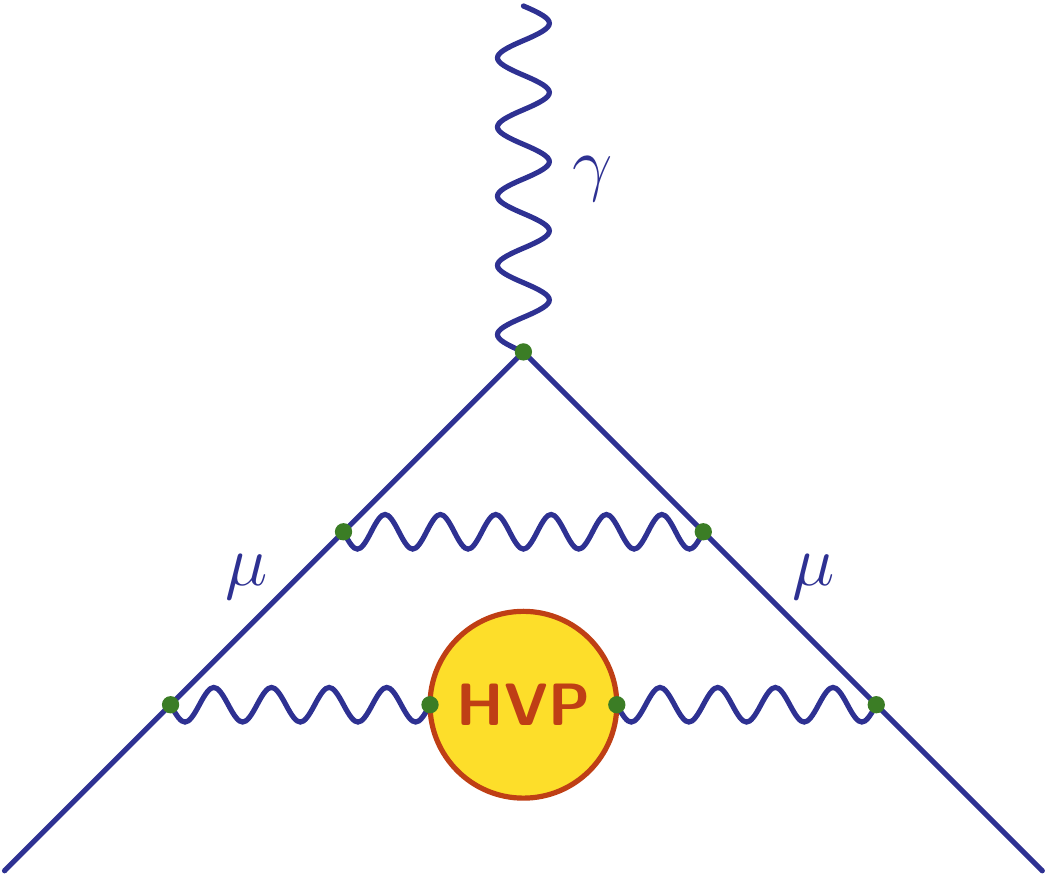} & &
\includegraphics[height=50mm,clip]{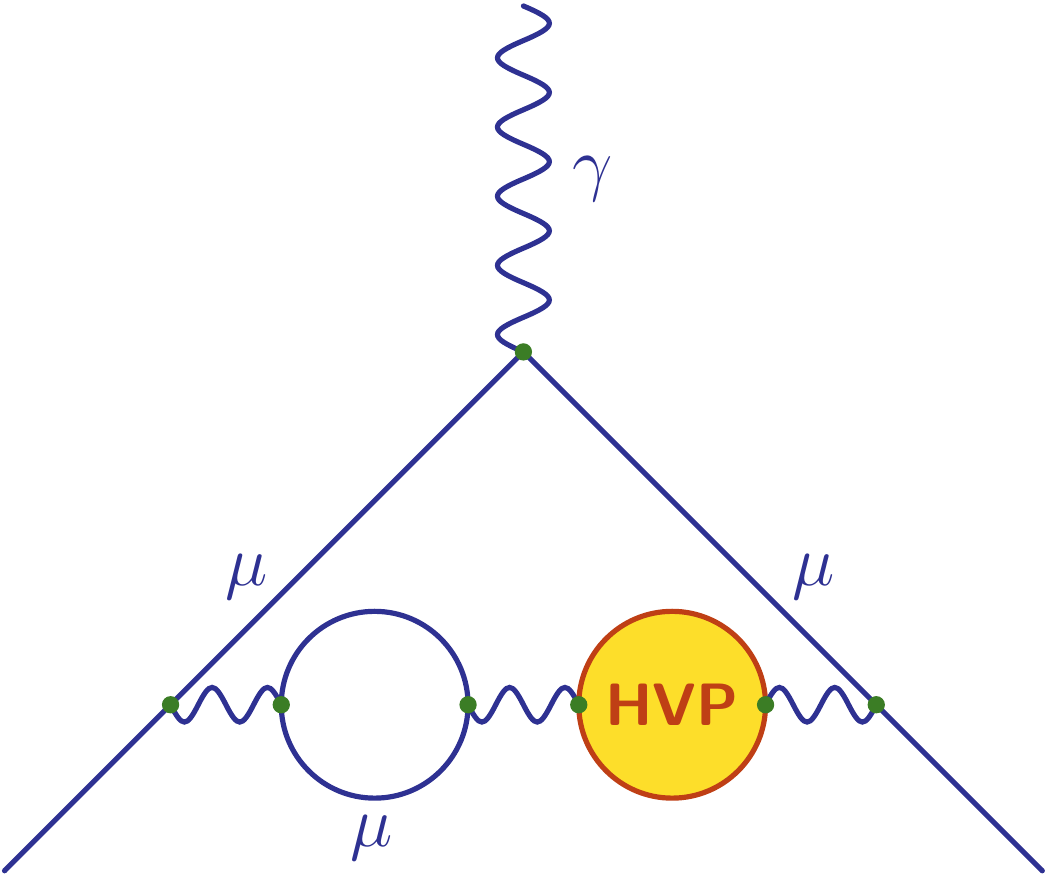} \\ 
\end{tabular}}
\caption{Two of the diagrams contributing to~$\amu{3a}$~(\ref{Amu3aDef}).}
\label{Plot:Amu3a}
\end{figure}

In~the next--to--leading order of perturbation theory
(namely, in the third order in the electromagnetic coupling)
the hadronic vacuum polarization contribution to
the muon anomalous magnetic moment~$\amu{3}$ is composed 
of three parts. Specifically, the first part~$\amu{3a}$ corresponds 
to the diagrams, which include (in~addition to the hadronic insertion
shown in~Fig.~\ref{Plot:Amu2}) one photon line or closed muon 
loop, see~Fig.~\ref{Plot:Amu3a}. In~turn, the second part~$\amu{3b}$ 
corresponds to the diagrams, which additionally include one closed 
electron (or~\mbox{$\tau$--lepton}) loop, whereas the 
third part~$\amu{3c}$ corresponds to the diagram with double
hadronic insertion. In what follows we shall primarily focus on the 
first part of~$\amu{3}$, which can be represented~as
\begin{equation}
\label{Amu3aDef}
\amu{3a} = \frac{2}{3} \Bigl(\frac{\alpha}{\pi}\Bigr)^{\!3}
\!\int\limits_{s_{0}}^{\infty}
\frac{G_{3a}(s)}{s} R(s) ds.
\end{equation}
The ``timelike'' kernel function~$G_{3a}(s)$ entering this equation
has been calculated explicitly in Ref.~\cite{DispMeth5}, 
see~App.~\ref{Sect:KR3aExpl}. However, the explicit form of the
corresponding kernel functions required for the assessment 
of~$\amu{3a}$ within ``spacelike'' methods is still unavailable.

\section{Results and discussion}
\label{Sect:Results}

\subsection{Relations between the kernel functions}
\label{Sect:Rels}

First of all, for practical purposes it is convenient to represent the
hadronic vacuum polarization contribution to the muon
anomalous magnetic moment, which corresponds to the
\mbox{$\ell$--th}~order in the electromagnetic coupling,
in the following form
\begin{subequations}
\label{Amu}
\begin{align}
\label{AmuP}
\amu{\ell} & = \APF{\ell}\!\!\int\limits_{0}^{\infty}\!
\KG{\Pi}{\ell}(Q^2) \bar\Pi(Q^2) \frac{d Q^2}{4\mmu^2} =
\APF{\ell}\!\!\int\limits_{0}^{\infty}\!
\KGt{\Pi}{\ell}(\zeta) \bar\Pi(4\zeta\mmu^2) d \zeta =
\\[1.25mm]
\label{AmuD}
& = \APF{\ell}\!\!\int\limits_{0}^{\infty} \!
\KG{D}{\ell}(Q^2) D(Q^2) \frac{d Q^2}{4\mmu^2} =
\APF{\ell}\!\!\int\limits_{0}^{\infty} \!
\KGt{D}{\ell}(\zeta) D(4\zeta\mmu^2) d \zeta = \\[1.25mm]
\label{AmuR}
& = \APF{\ell}\!\!\int\limits_{s_{0}}^{\infty} \!
\KG{R}{\ell}(s) R(s) \frac{d s}{4\mmu^2} =
\APF{\ell}\!\!\int\limits_{\chi}^{\infty} \!
\KGt{R}{\ell}(\eta) R(4\eta\mmu^2) d \eta.
\end{align}
\end{subequations}
In~this equation~$\APF{\ell}$ is a constant prefactor, 
$\bar\Pi(Q^2)=\DP(0,-Q^2)$ is~defined in~Eq.~(\ref{PSubDef}),
$Q^2 = -q^2 \ge 0$ and~$s = q^2 \ge 0$ stand, respectively,
for the spacelike and timelike kinematic variables, 
$\zeta = Q^2/(4\mmu^2)$ and~$\eta = s/(4\mmu^2)$ denote
the dimensionless kinematic variables, and~\mbox{$\chi=s_{0}/(4\mmu^2)$}. 
For~example, for the leading--order hadronic vacuum polarization
contribution~(\ref{Amu2Def})
\begin{equation}
\label{KR2eta}
\APF{2} = \frac{1}{3} \Bigl(\frac{\alpha}{\pi}\Bigr)^{\!2},
\qquad
\KGt{R}{2}(\eta) = \KG{R}{2}(4\eta\mmu^2) = 
G_{2}(4\eta\mmu^2) \frac{1}{\eta}.
\end{equation}

\begin{figure}[t]
\centerline{\includegraphics[width=75mm,clip]{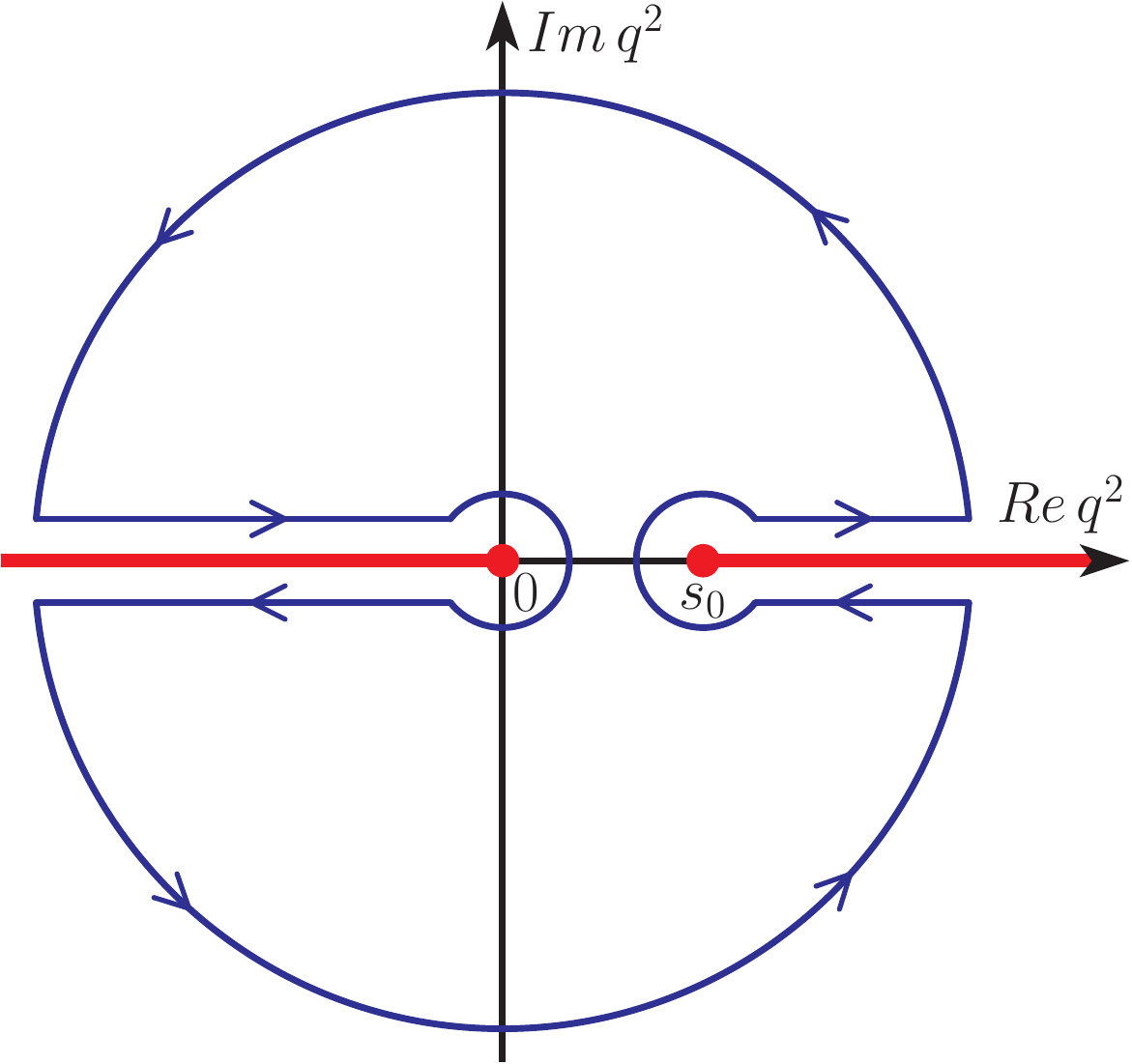}}
\caption{The closed integration contour~$C$ in the complex~$q^2$--plane
in Eq.~(\ref{IntC1}). The physical cut~$q^2 \ge s_{0}$ of 
the hadronic vacuum polarization 
function~$\Pi(q^2)=-\bar\Pi(-q^2)$~(\ref{PSubDef}) 
is shown along the positive semiaxis of real~$q^2$,
whereas the physical cut~\mbox{$q^2 \le 0$} of
the ``timelike'' kernel function~$K_{R}(q^2)$~(\ref{AmuR})
is shown along the negative
semiaxis of real~$q^2$.}
\label{Plot:Contour1}
\end{figure}

In~fact, the kernel functions~$K_{\Pi}(Q^2)$, $K_{D}(Q^2)$, 
and~$K_{R}(s)$ appearing in Eq.~(\ref{Amu}) can all be 
expressed in terms of each other. Let us begin by
expressing the ``spacelike'' kernel 
function~$K_{\Pi}(Q^2)$~(\ref{AmuP})
in terms of the ``timelike'' one~$K_{R}(s)$~(\ref{AmuR}).
As mentioned earlier, the hadronic vacuum polarization 
function~$\Pi(q^2)=-\bar\Pi(-q^2)$~(\ref{PSubDef}) 
possesses the only cut along the positive semiaxis
of real~$q^2$ starting at the hadronic production 
threshold~$q^2 \ge s_{0}$, whereas the kernel function
$K_{R}(q^2)$~(\ref{AmuR}) has the only cut along the negative 
semiaxis of real~$q^2$ starting at the origin~$q^2 \le 0$,
see, e.g., Ref.~\cite{DispMeth5}. Therefore, the integral of
the product of the functions~$K_{R}(q^2)$~(\ref{AmuR})
and~$\bar\Pi(-q^2)$~(\ref{PSubDef}) along
the contour displayed in Fig.~\ref{Plot:Contour1} vanishes, 
namely
\begin{equation}
\label{IntC1}
\oint_{C} K_{R}(q^2) \bar\Pi(-q^2) d q^2 = 0,
\end{equation}
that can also be represented~as
\begin{align}
\label{IntC2}
& \int\limits_{\infty-i\varepsilon}^{s_{0}-i\varepsilon} 
K_{R}(q^2) \bar\Pi(-q^2) d q^2 
+ \int\limits_{s_{0}+i\varepsilon}^{\infty+i\varepsilon}
K_{R}(q^2) \bar\Pi(-q^2) d q^2 
+ \nonumber \\ &
+ \int\limits_{-\infty+i\varepsilon}^{i\varepsilon} 
K_{R}(q^2) \bar\Pi(-q^2) d q^2 
+ \int\limits_{-i\varepsilon}^{-\infty-i\varepsilon}
K_{R}(q^2) \bar\Pi(-q^2) d q^2 = 0.
\end{align}
The change of the integration variables~$q^2 = p^2 - i\varepsilon$
in the first and fourth terms of Eq.~(\ref{IntC2}) 
and~$q^2 = p^2 + i\varepsilon$ in its second and third terms
casts Eq.~(\ref{IntC2})~to
\begin{equation}
\label{IntC3}
- \frac{1}{2 \pi i} \lim_{\varepsilon \to 0_{+}}
\int\limits_{0}^{-\infty} \bar\Pi(-p^2) 
\Bigl[ K_{R}(p^2+i\varepsilon) - K_{R}(p^2-i\varepsilon)\Bigr]
d p^2 =
\int\limits_{s_{0}}^{\infty} K_{R}(p^2) R(p^2) d p^2,
\end{equation}
where the limit~$\varepsilon \to 0_{+}$ 
is assumed and Eq.~(\ref{RDefP}) is employed.
Then the change of the integration variables~$p^2=-Q^2$
on the left--hand side of Eq.~(\ref{IntC3}) and~$p^2=s$
on its right--hand side leads~to
\begin{equation}
\label{IntC4}
\int\limits_{0}^{\infty} \bar\Pi(Q^2) K_{\Pi}(Q^2) d Q^2 =
\int\limits_{s_{0}}^{\infty} K_{R}(s) R(s) d s,
\end{equation}
where
\begin{equation}
\label{KRelPR}
K_{\Pi}(Q^2) = \frac{1}{2 \pi i} \lim_{\varepsilon \to 0_{+}}
\Bigl[ K_{R}(-Q^2+i\varepsilon) - K_{R}(-Q^2-i\varepsilon)\Bigr],
\qquad
Q^2 \ge 0.
\end{equation}
This relation has also been independently derived in a different way
in~Ref.~\cite{BLP}.

In turn, the relation inverse to Eq.~(\ref{KRelPR}) directly follows
from Eqs.~(\ref{AmuP}) and~(\ref{PDisp}), specifically
\begin{equation}
\int\limits_{0}^{\infty} 
K_{\Pi}(Q^2) \bar\Pi(Q^2) \frac{d Q^2}{4\mmu^2} =
\int\limits_{0}^{\infty} \frac{d Q^2}{4\mmu^2}
K_{\Pi}(Q^2)\, Q^2\!\! \int\limits_{s_{0}}^{\infty}
\frac{d s}{s}\, \frac{R(s)}{s+Q^2} =
\int\limits_{s_{0}}^{\infty} 
K_{R}(s) R(s) \frac{d s}{4\mmu^2},
\end{equation}
where
\begin{equation}
\label{KRelRP}
K_{R}(s) = \frac{1}{s} \int\limits_{0}^{\infty} 
K_{\Pi}(Q^2) \frac{Q^2}{s + Q^2}\, d Q^2,
\qquad
s \ge 0.
\end{equation}

The ``timelike'' kernel function~$K_{R}(s)$~(\ref{AmuR}) can be 
expressed in terms of the ``spacelike'' one~$K_{D}(Q^2)$~(\ref{AmuD})
in a similar way. In~particular, Eqs.~(\ref{AmuD}) and~(\ref{GDR_DR}) 
imply
\begin{equation}
\int\limits_{0}^{\infty} 
K_{D}(Q^2) D(Q^2) \frac{d Q^2}{4\mmu^2} =
\int\limits_{0}^{\infty} \frac{d Q^2}{4\mmu^2}
K_{D}(Q^2)\, Q^2\!\! \int\limits_{s_{0}}^{\infty}
\frac{R(s)}{(s+Q^2)^2}\, d s =
\int\limits_{s_{0}}^{\infty} 
K_{R}(s) R(s) \frac{d s}{4\mmu^2},
\end{equation}
where
\begin{equation}
\label{KRelRD}
K_{R}(s) = \int\limits_{0}^{\infty} 
K_{D}(Q^2) \frac{Q^2}{(s + Q^2)^2}\, d Q^2,
\qquad
s \ge 0.
\end{equation}

In turn, the corresponding relation between the kernel 
functions~$K_{\Pi}(Q^2)$~(\ref{AmuP})
and $K_{D}(Q^2)$~(\ref{AmuD}) can be derived from Eqs.~(\ref{GDR_DP}) 
and~(\ref{AmuD}), namely
\begin{align}
\label{KRelPDaux1}
& \int\limits_{0}^{\infty} 
K_{D}(Q^2) D(Q^2) d Q^2 =
-\int\limits_{0}^{\infty} d Q^2
K_{D}(Q^2)\, Q^2 \,\frac{d\,\Pi(-Q^2)}{d\, Q^2} =
\nonumber \\[1.25mm] &
= K_{D}(Q^2)\, Q^2 \,\bar\Pi(Q^2) \Biggr|_{0}^{\infty}
- \int\limits_{0}^{\infty} \bar\Pi(Q^2)
\biggl[K_{D}(Q^2) + \frac{d\,K_{D}(Q^2)}{d\,\ln Q^2}\biggr] d Q^2,
\end{align}
with the integration by parts being used. Since the first term of 
this equation vanishes (see also remarks given below), 
Eqs.~(\ref{AmuP}) and~(\ref{KRelPDaux1}) yield
\begin{equation}
\label{KRelPD}
K_{\Pi}(Q^2) = - \biggl[K_{D}(Q^2) 
+ \frac{d\,K_{D}(Q^2)}{d\,\ln Q^2}\biggr],
\qquad
Q^2 \ge 0.
\end{equation}

\begin{figure}[t]
\centerline{\includegraphics[width=75mm,clip]{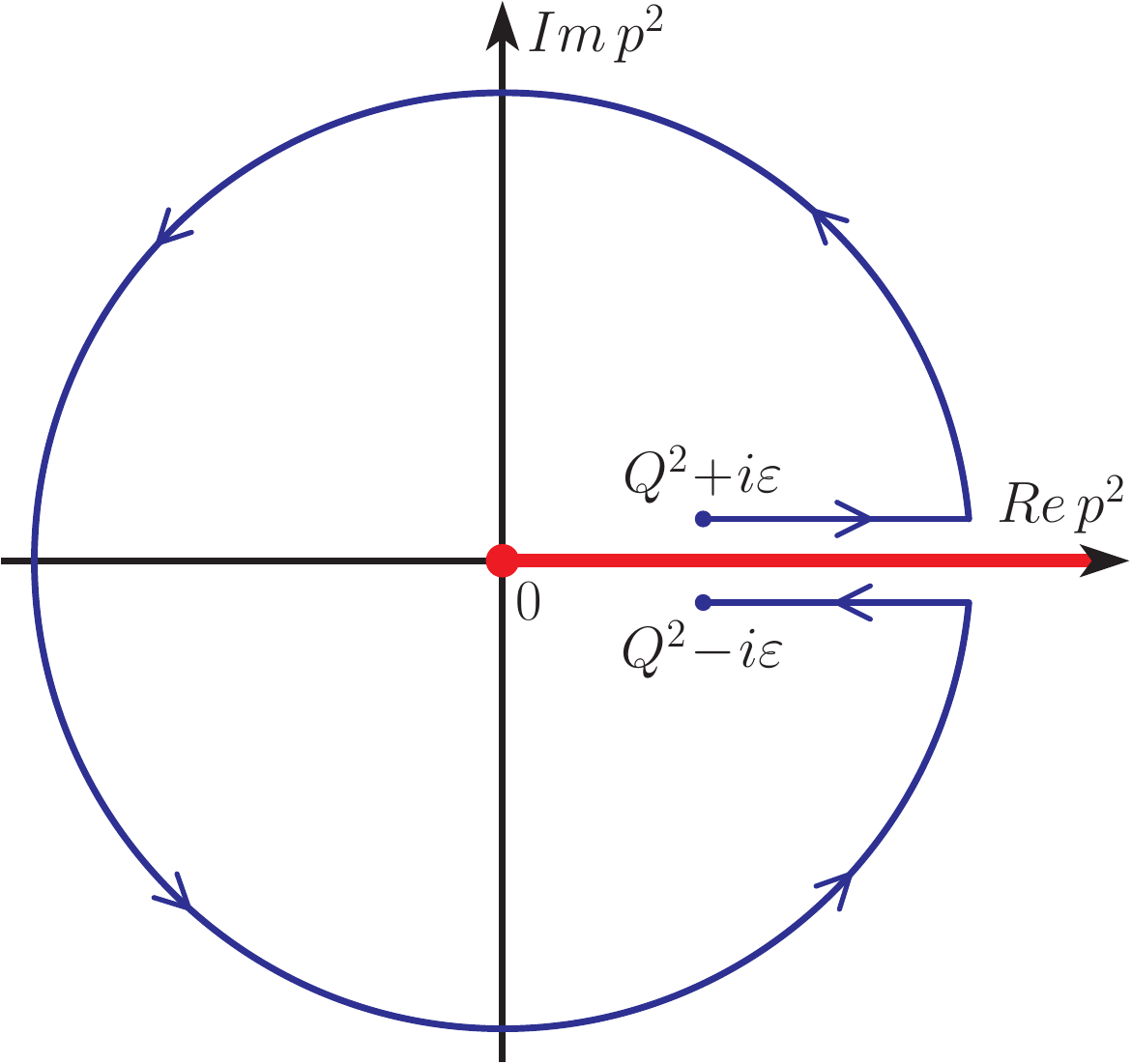}}
\caption{The integration contour in the complex~$p^2$--plane
in Eq.~(\ref{KRelDR}). The physical cut $p^2 \ge 0$ of the 
``timelike'' kernel function~$K_{R}(-p^2)$~(\ref{AmuR}) is shown 
along the positive semiaxis of real~$p^2$.}
\label{Plot:Contour2}
\end{figure}

The kernel function~$K_{D}(Q^2)$~(\ref{AmuD}) can be expressed in terms
of~$K_{\Pi}(Q^2)$~(\ref{AmuP}) in the following way. The solution to the 
differential equation~(\ref{KRelPD})
\begin{equation}
\label{KRelDPaux1}
K_{D}(Q^2) + \frac{d\,K_{D}(Q^2)}{d\,\ln Q^2} = - K_{\Pi}(Q^2)
\end{equation}
reads
\begin{equation}
\label{KRelDPaux2}
K_{D}(Q^2) = \frac{1}{Q^2} 
\biggl[-\!\int\!\! K_{\Pi}(Q^2)\, d Q^2 + c_{0} \biggr]\!,
\end{equation}
where~$c_{0}$ denotes an arbitrary integration constant.
The latter has to be chosen in the way that prevents the
appearance of the mutually canceling divergences at the
upper limit of both terms in the second line of 
Eq.~(\ref{KRelPDaux1}). Specifically, the integration 
constant~$c_{0}$ has to subtract the value of the
antiderivative of the function~$K_{\Pi}(Q^2)$ 
at~$Q^2 \to \infty$, that eventually results~in
\begin{equation}
\label{KRelDP}
K_{D}(Q^2) = \frac{1}{Q^2} 
\int\limits_{Q^2}^{\infty}\!\! K_{\Pi}(\xi)\, d \xi,
\qquad
\xi = -p^2 \ge 0,
\end{equation}
where~$\xi$ stands for a spacelike kinematic variable.
It~is worthwhile to note also that Eqs.~(\ref{KRelDP})
and~(\ref{KRelRP}) imply that the functions~$Q^2 K_{D}(Q^2)$ 
and~$sK_{R}(s)$ acquire the same value in the infrared limit, 
namely
\begin{equation}
\label{KRDlim}
\lim_{Q^2 \to 0_{+}} Q^2 K_{D}(Q^2) =
\lim_{s \to 0_{+}} sK_{R}(s) =
\int\limits_{0}^{\infty}\!\! K_{\Pi}(\xi)\, d \xi.
\end{equation}

Finally, the relation inverse to Eq.~(\ref{KRelRD}) can be derived
by making use of Eqs.~(\ref{KRelDP}) and~(\ref{KRelPR}), namely
\begin{equation}
\label{KRelDRaux1}
K_{D}(Q^2) = - \frac{1}{2 \pi i} \lim_{\varepsilon \to 0_{+}}
\frac{1}{Q^2} 
\int\limits_{Q^2}^{\infty}
\Bigl[ K_{R}(-\xi-\iep) - K_{R}(-\xi+\iep)\Bigr] d \xi.
\end{equation}
The change of the integration variables~$\xi = p^2 - \iep$
and~$\xi = p^2 + \iep$ in, respectively, the first and the 
second terms in the square brackets in Eq.~(\ref{KRelDRaux1}) 
eventually leads~to
\begin{equation}
\label{KRelDR}
K_{D}(Q^2) = - \frac{1}{2 \pi i} \lim_{\varepsilon \to 0_{+}}
\frac{1}{Q^2} 
\int\limits_{Q^2 + \iep}^{Q^2 - \iep}
K_{R}(-p^2) d p^2,
\end{equation}
where the integration contour on the right--hand side of this equation 
lies in the region of analyticity of the function~$K_{R}(-p^2)$, 
see Fig.~\ref{Plot:Contour2}.

Thus, the derived equations~(\ref{KRelPR}), (\ref{KRelRP}), 
(\ref{KRelRD}), (\ref{KRelPD}), (\ref{KRelDP}), and~(\ref{KRelDR}) 
constitute the complete set of relations, which mutually express 
the ``spacelike'' and ``timelike'' kernel 
functions~$K_{\Pi}(Q^2)$, $K_{D}(Q^2)$, and~$K_{R}(s)$ 
entering Eq.~(\ref{Amu}) in terms of each other.

\subsection{The ``spacelike'' kernel functions}
\label{Sect:KP3aExpl}

The relations obtained in Sect.~\ref{Sect:Rels} enable one to
calculate the unknown kernel functions~(\ref{Amu})
by making use of the known ones. To~exemplify this method,
let us first address the hadronic vacuum polarization contribution 
to the muon anomalous magnetic moment in the leading~order.

Specifically, the explicit form of the ``spacelike'' kernel 
function~$\KG{\Pi}{2}(Q^2)$~(\ref{AmuP}) can be obtained directly
from the corresponding ``timelike'' 
one~$\KG{R}{2}(s)$ [Eqs.~(\ref{K2RExpl}),~(\ref{KR2eta})]
by making use of the derived relation~(\ref{KRelPR}),
that eventually results~in
\begin{equation}
\label{KP2expl}
\zeta\KGt{\Pi}{2}(\zeta) = \frac{1}{\zeta^2} \, 
\frac{y^{5}(\zeta)}{1-y(\zeta)},
\qquad
y(\zeta) = \zeta \Bigl(\sqrt{1+\zeta^{-1}}-1\Bigr),
\qquad
\zeta = \frac{Q^2}{4\mmu^2}.
\end{equation}
It is worth noting that Eq.~(\ref{KP2expl}) is identical 
to the result of the mapping the integration range~$0 \le x < 1$
in~Eq.~(\ref{Amu2Px}) onto the kinematic 
interval~\mbox{$0 \le Q^2 < \infty$}
reported~in Refs.~\cite{Pivovarov2002, JPG42, EdR17}
and to the result of straightforward calculation~\cite{Blum2003}
performed within the technique~\cite{HS1, HS2, HS3, HS4}.

In turn, the explicit form of the kernel 
function~$\KG{D}{2}(Q^2)$~(\ref{AmuD}) can be derived 
from Eq.~(\ref{KP2expl}) by making use of the obtained 
relation~(\ref{KRelDP}), that leads~to
\begin{equation}
\label{KD2expl}
\zeta\KGt{D}{2}(\zeta) = (2\zeta+1)^{2} 
- 2(2\zeta+1)\sqrt{\zeta(\zeta+1)} - \frac{1}{2}.
\end{equation}
This equation coincides with the result of the mapping the 
integration range~$0 \le x < 1$ in~Eq.~(\ref{Amu2Dx}) onto 
the kinematic interval~\mbox{$0 \le Q^2 < \infty$}, 
see~Refs.~\cite{Pivovarov2002, EdR17}.

\begin{figure}[t]
\centerline{\includegraphics[width=120mm,clip]{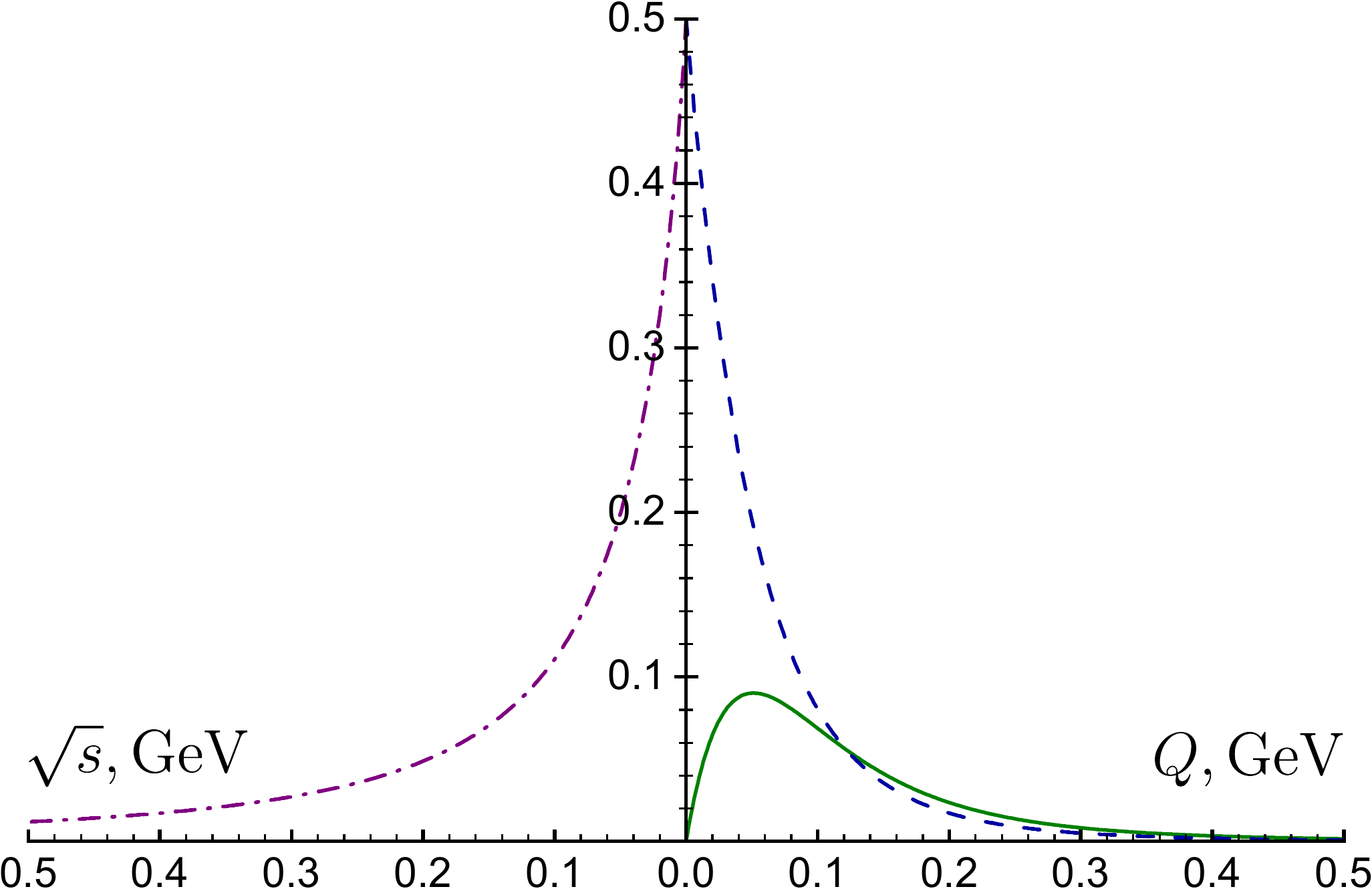}}
\caption{The kernel functions
$\zeta\KGt{\Pi}{2}(\zeta)$ [Eq.~(\ref{KP2expl}), solid curve], 
$\zeta\KGt{D}{2}(\zeta)$ [Eq.~(\ref{KD2expl}), dashed curve],
and 
$\eta\KGt{R}{2}(\eta)$ [Eqs.~(\ref{K2RExpl}),~(\ref{KR2eta}), 
dot--dashed curve]
in the 
spacelike [$Q^2 = -q^2 \ge 0$, \mbox{$\zeta = Q^2/(4\mmu^2)$}]
and timelike [$s = q^2 \ge 0$, $\eta = s/(4\mmu^2)$]
domains.}
\label{Plot:K2PRD}
\end{figure}

The ``spacelike'' [$\KG{\Pi}{2}(Q^2)$, Eq.~(\ref{KP2expl})
and $\KG{D}{2}(Q^2)$, Eq.~(\ref{KD2expl})] 
and ``timelike'' [$\KG{R}{2}(s)$, Eqs.~(\ref{K2RExpl}),~(\ref{KR2eta})]
kernel functions
satisfy all six relations~(\ref{KRelPR}), (\ref{KRelRP}), 
(\ref{KRelRD}), (\ref{KRelPD}), (\ref{KRelDP}), 
and~(\ref{KRelDR}) derived in Sect.~\ref{Sect:Rels}.
The plots of the kernel functions (\ref{KP2expl}), 
(\ref{KD2expl}), and~(\ref{K2RExpl}) are displayed
in~Fig.~\ref{Plot:K2PRD}. In particular, as~one can 
infer from this figure, in~the infrared limit
the kernel functions 
$\zeta\KGt{D}{2}(\zeta)$~(\ref{KD2expl})
and
$\eta\KGt{R}{2}(\eta)$~(\ref{K2RExpl}) 
assume the same value~$1/2$, as
determined by the relation~(\ref{KRDlim}).

\bigskip

In the next--to--leading~order the explicit form of the uncalculated 
yet ``spacelike'' kernel function $\KG{\Pi}{3a}(Q^2)$ appearing in 
Eq.~(\ref{AmuP}) can be obtained in a similar way. Specifically,
the derived relation~(\ref{KRelPR}) and 
Eq.~(\ref{KR3aExpl}) eventually result~in
\begin{align}
\label{KP3aExpl}
\zeta \tilde{K}\ind{(3a)}{$\Pi$}(\zeta) & = 
- \biggl[
  \frac{19}{12} + \frac{7}{9}\zeta 
+ \frac{23}{9}\zeta^2 - \frac{1}{4(\zeta+1)} 
  \biggr] 
+ \nonumber \\[1mm] & 
+ \biggl(
  \frac{1}{3\zeta} + \frac{127}{36} 
+ \frac{115}{18}\zeta + \frac{23}{9}\zeta^2 
  \biggr) \psi(\zeta+1)
- \nonumber \\[1mm] & 
- \frac{5}{3} \zeta^2 \ln(4\zeta)
- \biggl( \frac{14}{3}\zeta + 1 \biggr)
  (\zeta+1)
  \psi(\zeta+1) 
\times \nonumber \\[1mm] & 
\times 
  \biggl\{
  \frac{1}{2}\ln(4\zeta) + 3A(\zeta+1) + 2\ln\Bigl[1+B(\zeta+1)\Bigr]\!
  \biggr\} 
+ \nonumber \\[1mm] & 
+ \biggl( 
  -\frac{19}{6} + \frac{53}{3}\zeta + \frac{58}{3}\zeta^2
  -\frac{1}{3\zeta} + \frac{2}{\zeta+1}  
  \biggr)
  A(\zeta+1)
+ \nonumber \\[1mm] & 
+ \biggl[ 
  \frac{13}{12\,\zeta} + \frac{7}{6} + \zeta
  +\frac{8}{3}\zeta^2 + \frac{1}{4\zeta(\zeta+1)}
  \biggr]
  \psi(\zeta+1)
  A(\zeta+1) 
- \nonumber \\[1mm] & 
- \biggl(
  \frac{1}{2} + \frac{14}{3}\zeta + 8\zeta^2
  \biggr)
\times \nonumber \\[1mm] & 
\times 
  \biggl\{
  2A(\zeta+1)
  \Bigl\{
  2\ln\bigl[1+B(\zeta+1)\bigr] + \ln\bigl[1-B(\zeta+1)\bigr]\!
  \Bigr\}
  - 
\nonumber \\[1mm] & 
  -2\Bigl\{
  \Li{2}\bigl[B(\zeta+1)\bigr] 
  +2\Li{2}\bigl[-B(\zeta+1)\bigr]\!
  \Bigr\}\!
  \biggr\}.
\end{align}
An~equivalent form of this equation has been independently 
derived in~Ref.~\cite{BLP}. In~Eq.~(\ref{KP3aExpl})
$\zeta=Q^2/(4\mmu^2)$,
$Q^2 = -q^2 \ge 0$ stands for the spacelike kinematic
variable, the functions~$\psi(\zeta)$ and~$A(\zeta)$
are defined in~Eq.~(\ref{DefAux1}),
\begin{equation}
\label{DefAux2}
B(\zeta) = \frac{1-\psi(\zeta)}{1+\psi(\zeta)},
\end{equation}
and
\begin{equation}
\label{Li2Def}
\Li{2}(y) = -\int\limits_{0}^{y}\ln(1-t)\frac{d\,t}{t}
\end{equation}
denotes the dilogarithm function. 

\begin{figure}[t]
\centerline{\includegraphics[width=120mm,clip]{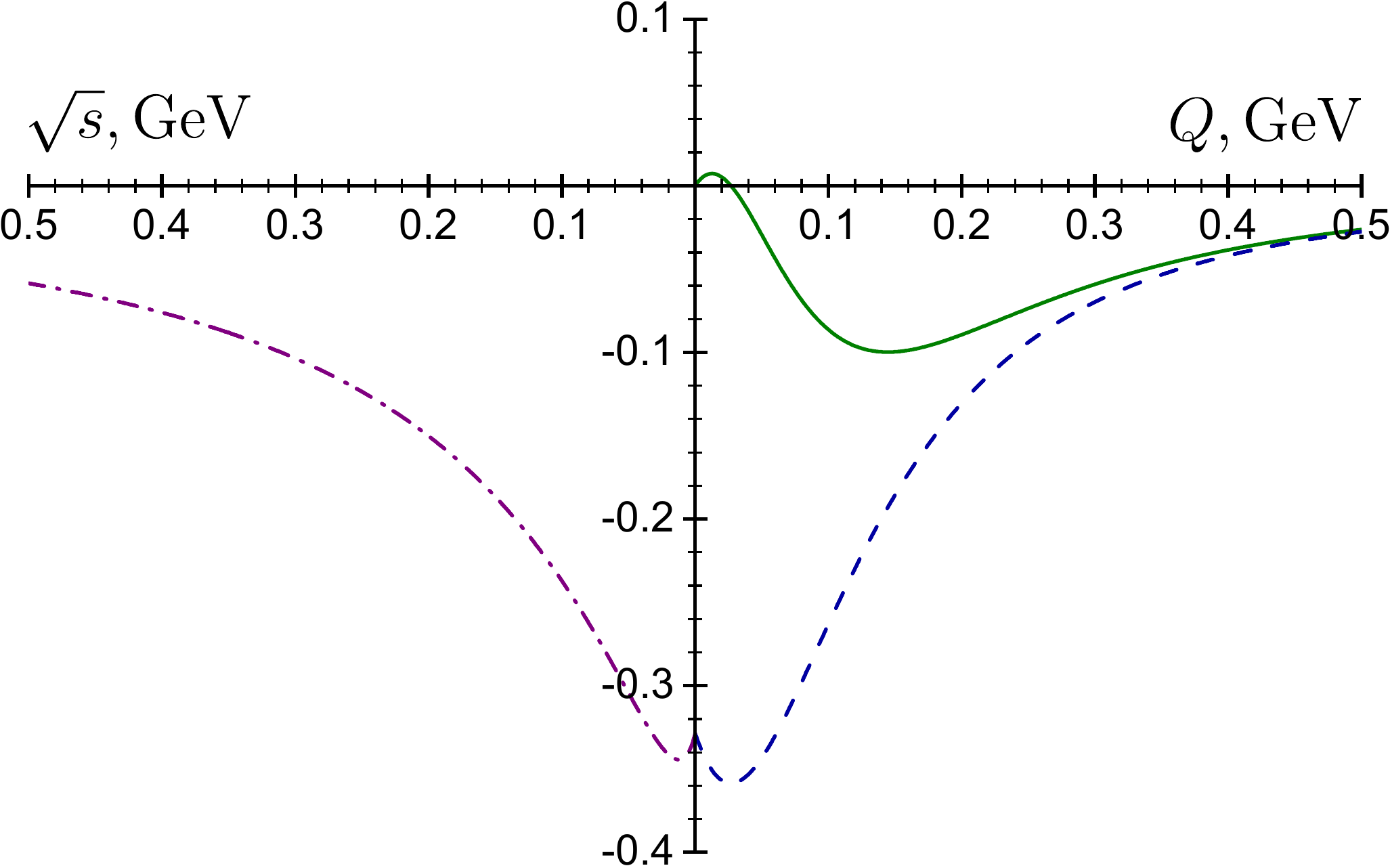}}
\caption{The kernel functions
$\zeta\KGt{\Pi}{3a}(\zeta)$ [Eq.~(\ref{KP3aExpl}), solid curve], 
$\zeta\KGt{D}{3a}(\zeta)$ [Eqs.~(\ref{KRelDP}),~(\ref{KP3aExpl}), 
dashed curve],
and 
$\eta\KGt{R}{3a}(\eta)$ [Eq.~(\ref{KR3aExpl}), dot--dashed curve]
in the 
spacelike [$Q^2 = -q^2 \ge 0$, \mbox{$\zeta = Q^2/(4\mmu^2)$}]
and timelike [$s = q^2 \ge 0$, $\eta = s/(4\mmu^2)$]
domains.}
\label{Plot:K3aPR}
\end{figure}

It is straightforward to verify that the 
``spacelike'' [$\KG{\Pi}{3a}(Q^2)$, Eq.~(\ref{KP3aExpl})] 
and ``timelike'' [$\KG{R}{3a}(s)$, Eq.~(\ref{KR3aExpl})]
kernel functions satisfy the corresponding 
relations~(\ref{KRelPR}) 
and~(\ref{KRelRP}) obtained in Sect.~\ref{Sect:Rels}.
The~plots of the kernel 
functions~[$\KG{\Pi}{3a}(Q^2)$, Eq.~(\ref{KP3aExpl})], 
[$\KG{D}{3a}(Q^2)$, computed numerically by making use
of~Eqs.~(\ref{KRelDP}) and~(\ref{KP3aExpl})],
and~[$\KG{R}{3a}(s)$, Eq.~(\ref{KR3aExpl})] 
are displayed in~Fig.~\ref{Plot:K3aPR}.
As~one can infer from this figure, in~the infrared limit
the kernel functions 
$\zeta\KGt{D}{3a}(\zeta)$~[Eqs.~(\ref{KRelDP}),~(\ref{KP3aExpl})]
and
$\eta\KGt{R}{3a}(\eta)$~[Eq.~(\ref{KR3aExpl})] acquire the same 
value determined by the relation~(\ref{KRDlim}), specifically
\begin{equation}
\lim_{\zeta \to 0_{+}} \zeta\KGt{D}{3a}(\zeta) =
\lim_{\eta \to 0_{+}} \eta\KGt{R}{3a}(\eta) =
\frac{197}{144} +\frac{1}{2}\zeta_{2} 
-3\zeta_{2}\ln(2) +\frac{3}{4}\zeta_{3} 
\simeq -0.328479,
\end{equation}
where
\begin{equation}
\label{RiemannZeta}
\zeta_{t} = \sum_{n=1}^{\infty} \frac{1}{n^{t}}
\end{equation}
stands for the Riemann $\zeta$~function.

\section{Conclusions}
\label{Sect:Concl}

The complete set of relations [Eqs.~(\ref{KRelPR}), (\ref{KRelRP}), 
(\ref{KRelRD}), (\ref{KRelPD}), (\ref{KRelDP}),~(\ref{KRelDR})], 
which mutually express the~``spacelike'' [$K_{\Pi}(Q^2)$, Eq.~(\ref{AmuP}) 
and $K_{D}(Q^2)$, Eq.~(\ref{AmuD})] and~``timelike'' [$K_{R}(s)$, 
Eq.~(\ref{AmuR})] kernel functions in terms of each other, is obtained. 
By~making use of the derived relations the explicit expression for the
next--to--leading order ``spacelike'' kernel function~$\KG{\Pi}{3a}(Q^2)$ 
is calculated [Eq.~(\ref{KP3aExpl})] and the kernel 
function~$\KG{D}{3a}(Q^2)$ is computed numerically 
[Eqs.~(\ref{KRelDP}),~(\ref{KP3aExpl}) and~Fig.~\ref{Plot:K3aPR}]. 
The~obtained results can be employed in the assessments 
of the hadronic vacuum polarization contributions to the muon
anomalous magnetic moment in the framework of the spacelike methods,
such as lattice studies~\cite{Lattice1, Lattice2}, MUonE 
project~\cite{MUonE1, MUonE2, MUonE3}, and others.

\appendix

\section{The ``timelike'' kernel function~$K\ind{(3a)}{$R$}(s)$}
\label{Sect:KR3aExpl}

As mentioned earlier, the next--to--leading 
order hadronic vacuum polarization contribution to the muon 
anomalous magnetic moment~(\ref{Amu3aDef}) can be represented~as
\begin{equation}
\label{Amu3aAux}
\amu{3a} = \APF{3a}\!\!\int\limits_{s_{0}}^{\infty} \!
\KG{R}{3a}(s) R(s) \frac{d s}{4\mmu^2} =
\APF{3a}\!\!\int\limits_{\chi}^{\infty} \!
\KGt{R}{3a}(\eta) R(4\eta\mmu^2) d \eta,
\end{equation}
where
\begin{equation}
\APF{3a} = \frac{2}{3} \Bigl(\frac{\alpha}{\pi}\Bigr)^{\!3},
\quad
\KGt{R}{3a}(\eta) = \KG{R}{3a}(4\eta\mmu^2) = 
G_{3a}(4\eta\mmu^2) \frac{1}{\eta},
\quad
\eta=\frac{s}{4\mmu^2},
\quad
\chi=\frac{s_{0}}{4\mmu^2}.
\end{equation}
The explicit form of the ``timelike'' kernel 
function entering Eq.~(\ref{Amu3aAux}) was calculated
in~Ref.~\cite{DispMeth5}, namely
\begin{align}
\label{KR3aExpl}
\eta \tilde{K}\ind{(3a)}{$R$}(\eta) & = 
-\frac{139}{144} + \frac{115}{18}\eta 
+ \nonumber \\[1mm] & 
+ \biggl[
  \frac{19}{12} - \frac{7}{9}\eta 
  + \frac{23}{9}\eta^2 + \frac{1}{4(\eta-1)}
  \biggr] \ln(4\eta)
+ \nonumber \\[1mm] & 
+ \biggl[
  \frac{2}{3\,\eta} - \frac{127}{18} 
  + \frac{115}{9}\eta - \frac{46}{9}\eta^2
  \biggr]
  \frac{1}{\psi(\eta)}A(\eta)
+ \nonumber \\[1mm] & 
+ \biggl(
  \frac{9}{4} + \frac{5}{6}\eta 
  - 8\eta^2 - \frac{1}{2\,\eta} 
  \biggr) 
  \zeta_{2}
  + 
  \frac{5}{6}\eta^{2}\ln^{2}(4\eta)
+ \nonumber \\[1mm] & 
+ \biggl(
  \frac{14}{3}\eta -1
  \biggr)
  (\eta-1) \frac{1}{\psi(\eta)} T_{1}(\eta)
+ \nonumber \\[1mm] & 
+ \biggl(
  \frac{19}{6} + \frac{53}{3}\eta 
  - \frac{58}{3}\eta^2 - \frac{1}{3\,\eta} + \frac{2}{\eta-1}
  \biggr)
  A^{2}(\eta)
+ \nonumber \\[1mm] & 
+ \biggl[
  \frac{13}{12\,\eta} -\frac{7}{6} + \eta
  - \frac{8}{3}\eta^2 - \frac{1}{4\eta(\eta-1)}
  \biggr]
  \frac{1}{\psi(\eta)} T_{2}(\eta)
+ \nonumber \\[1mm] & 
+ \biggl(
  \frac{1}{2} - \frac{14}{3}\eta + 8\eta^2
  \biggl)
  T_{3}(\eta),
\end{align}
where~$\eta = s/(4\mmu^2)$, $s = q^2 \ge 0$ is the timelike kinematic 
variable, the functions~$\psi(\eta)$, $A(\eta)$,
and~$B(\eta)$ were given in Eqs.~(\ref{DefAux1}) 
and~(\ref{DefAux2}), respectively,
\begin{equation}
\label{T1Def}
T_{1}(\eta) = 
  A(\eta) \ln(4\eta)
+ 2 \biggl\{\!
  \Li{2}\Bigl[1-B(\eta)\Bigr] + A^{2}(\eta)
  \!\biggr\},
\end{equation}
\begin{equation}
\label{T2Def}
T_{2}(\eta) = 
  \Li{2}\Bigl[-B(\eta)\Bigr] + A^{2}(\eta) + \frac{1}{2}\zeta_{2},
\end{equation}
\begin{align}
\label{T3Def}
T_{3}(\eta) & = 
- 6\Li{3}\Bigl[B(\eta)\Bigr] 
- 3\Li{3}\Bigl[-B(\eta)\Bigr] 
+ 4\ln\Bigl[1-B(\eta)\Bigr] A^{2}(\eta)
+ \nonumber \\[1mm] & 
+ \Bigl[2A^{2}(\eta) + 3\zeta_{2}\Bigr] 
  \ln\Bigl[1+B(\eta)\Bigr]
- \nonumber \\[1mm] & 
- 4\biggl\{\!
  \Li{2}\Bigl[-B(\eta)\Bigr]
  + 2\Li{2}\Bigl[-B(\eta)\Bigr]
  \!\biggr\} A(\eta),
\end{align}
the dilogarithm 
and Riemann $\zeta$~functions
have been defined in Eqs.~(\ref{Li2Def}) and~(\ref{RiemannZeta}), 
respectively,~and
\begin{equation}
\label{Li3RZ}
\Li{3}(y) = \int\limits_{0}^{y}\Li{2}(t)\frac{d\,t}{t}
\end{equation}
denotes the trilogarithm~function.

\end{document}